# MOLECULAR STRUCTURE AND ELECTROPHYSICAL PROPERTIES OF PENTACENE THIODERIVATIVES

M.P. GORISHNYI

UDC 535.343.2;535.343.3;537.311.322;537.312.5

Institute of Physics, Nat. Acad. of Sci. of Ukraine

(46, Nauky Ave., Kyiv 03028, Ukraine; e-mail: gorishny@iop.kiev.ua)

Possible conformations of the thioderivatives of pentacene (Pn) have been considered. The absorption spectra of polythiopentacene (PTPn) solutions and films have been studied. PTPn is revealed to be a mixture of Pn thioderivatives with different numbers of S atoms. After this mixture having been condensed in vacuum onto quartz substrates, its main components are tetrathiopentacene (TTPn) and hexathiopentacene (HTPn). The position of the maximum in the long-wave absorption bands of Pn thioderivatives is a linear function of the number of valence electrons in S atoms, which take part in the conjugation with the π-system of the pentacene frame of PTPn molecules. The analysis of the photocurrent and capacitor photovoltage (CPV) spectra in the range of the first electron transitions in PTPn has shown that the photoconductivity is of the hole type and is caused by the dissociation of excitons at the electron capture centers. The frontal CPV is caused by the Dember photovoltage ($\varphi_D$), and the back one by the surface-barrier photovoltage ($\varphi_b$).

## 1. Introduction

The thioderivatives of linear polyacenes have been synthesized for the first time in [1, 2]. The scientific interest in studying their properties is stimulated by the rather low electric resistance of tetrathiotetracene (TTT), as compared to that of organic materials [3]. Later on, it has been discovered that tetraselenotetracene (TSeT) and hexathiohexacene (HTH) possess the resistance of the same order ($\rho(300\ K) = 10^2\ \Omega \times m$) [4].

TTT has been studied the best [5, 6]. This substance and its isostructural analog TSeT belong to strong electron donors. On their basis, ion-radical salts with high conductivity have been synthesized; some of the latter reveal the metal conductivity in a wide range of temperatures [7, 8]. Nevertheless, the efficiency of the conversion of light energy into electric one, making use of TTT-based thin-film heterostructures, is low, which is caused by the low photosensitivity of this compound [9].

Among the linear polyacenes, which have been studied, high-resistance pentacene reveals rather high photosensitivity [10]. It forms thioderivatives with different numbers of S atoms. The photovoltaic and optical properties of the mixture of those thioderivatives (hereafter, PTPn) have been studied in [11]. The most probable components of PTPn are TTPn and HTPn. The nature of their long-wave absorption bands has been established in [12]. The photovoltaic properties of HTPn-based thin-film heterostructures have been studied in [13]. HTPn served as a basis for fabricating the field transistors [14].

In work [13], the purity degree of HTPn under investigation has not been discussed, and its photovoltaic properties have not been studied enough. This problem is challenging from the viewpoint of searching for new photosensitive organic materials with relatively low dark electroconductivity. Thioderivatives of pentacene with different numbers of S atoms can be characterized by different values of this parameter.

In this work, an attempt has been made to isolate the PTPn components and to study the properties of this substance in more detail.

## 2. Synthesis of Pentacene Thioderivatives and Experimental Method

Pentacene thioderivatives were synthesized by Doctor L. Libera at the Supreme Polytechnic School in Chemnitz (Germany). The mixture of pentacene and sulfur was heated up in trichlorobenzene. The solution quickly became green. The reaction was supervised by the disappearance of the colors of initial substances. The crystals formed were dried in vacuum and washed out in alcohol. The chemical analysis of the final substance showed that the content of sulfur in it was 36.5%, i.e. an intermediate value between those in stoichiometric TTPn (31.6%) and HTPn (41.4%). Relying on these data, the substance obtained was identified as PTPn [12].

In the second case, the synthesis was carried out following the technique, according to which the substance obtained was additionally purified through its recrystallization from a solution in nitrobenzene. Chemical analysis showed that the content of sulfur in the final substance was close to that in stoichiometric HTPn.

PTPn and HTPn films were fabricated by thermal sputtering of the relevant substance in vacuum of $6\times10^{-4}$ Pa onto quartz substrates at room temperature. The film thickness was measured making use of an MII-4 microscope, while the film thickness variation in the course of sputtering and the sputtering rate were determined by an MSV-1841 thickness indicator.

Specimens were illuminated with a halogen incandescent lamp either directly or through an SPM-2 monochromator. The power of the incident fixed light was measured making use of a PTH-20C thermopile, and that of the incident modulated light by a pyrometer with sensitivities of 0.83 and 250 V/W, respectively. The resistance of surface cells in darkness and under illumination was determined with the help of a P4053 direct current bridge. These data, after recalculation, were used to obtain the photocurrent spectra.

The measurements of CPV was carried on in a capacitor cell, where PTPn and HTPn films were deposited onto a quartz substrate with a conducting layer of SnO2 and isolated from a similar substrate by a transparent dielectric Teflon film 10 μm in thickness. The cell capacitance was 150 – 200 pF. The cell was illuminated with light modulated with a frequency of 72 Hz.

If the photoresponse of the capacitor cell is sinusoidal, one may examine its equivalent electric circuits (series and parallel) in order to determine the effective input resistance and capacitance of the amplifier in the first cascade of the installation. This is a rough approximation, but it allows one to estimate the orders of magnitude of these parameters.

The $SnO_2$ electrodes were found quasi-Ohmic for Pn thioderivatives. In the first approximation, their resistances and capacitances are neglected. Then, the cell is a series circuit composed of the resistance of the PTPn or HTPn layer and the reactance of the capacitor that is formed by the free PTPn (HTPn) surface, the Teflon layer, and the $SnO_2$ electrode. For a 560-nm PTPn film, a 200-pF capacitor, and at the 72-Hz modulation frequency, the calculated values of these parameters for a cell are as follows: the resistance R = 50 MΩ, the reactance X = 11 MΩ, and the impedance Z = 51 MΩ. If the series circuit is replaced by a parallel one, the parameters of the latter will be functions of the cyclic frequency ω. Their calculated values, provided Z = const, are as follows: R(ω) = 52.5 MΩ and C (ω) = 9.3 pF. Therefore, the following relations are valid: R(ω) ≈ R and C (ω) ï C . The obtained values of R(ω) and C (ω) were taken into account while designing the amplifier block of the installation in order to ensure the regime of measuring the CPV.

A cathode follower with compensation of the input capacitance [15] was used as the input stage. The effective input resistance of the installation was not lower than 1 GΩ, with the input capacitance C ≤ 1 pF. After preamplification, the signal was registered by an UPI-1 selective phase-sensitive amplifier. The photocurrent and CPV spectra were recalculated to the identical power of the incident light.

The absorption spectra of PTPn and HTPn solutions and films were recorded making use of a Hitachi double-beam spectrometer, provided a spectral slit width of 2 nm and at room temperature.

3. Discussion of the Results Obtained

Possible structural formulae of the molecules of pentacene thioderivatives are summarized in Fig. 1. Their pentacene skeleton (five linearly arranged benzene rings) possesses six chemically active C atoms, designated by circles. Sulfur atoms can add to them. If the number of added S atoms is four, TTPn is formed; the latter can exist in two isomeric forms depending on a relative arrangement of dithiol groups (–S–S–). Fig. 1,a exposes only that TTPn isomer, in which dithiol groups are located against each other, as in TTT.

The π-system of the pentacene skeleton is formed by the overlapping of $2p_z$-orbitals of C atoms and is regarded as a benzene sextet delocalized along the external perimeter of five rings. The electronic configuration of the external electrons of the S atoms looks like $3S^2 3P^4$. According to such a structure, σ- bonds of the dithiol groups are formed by $3p_x$- and $3p_y$-orbitals. $3p_z$-orbitals of coupled electrons are orthogonal to the pentacene skeleton plane and may overlap with $2p_z$-orbitals of chemically active C atoms, so that the π- system of the TTPn molecule changes its own geometry and the number of delocalized electrons.

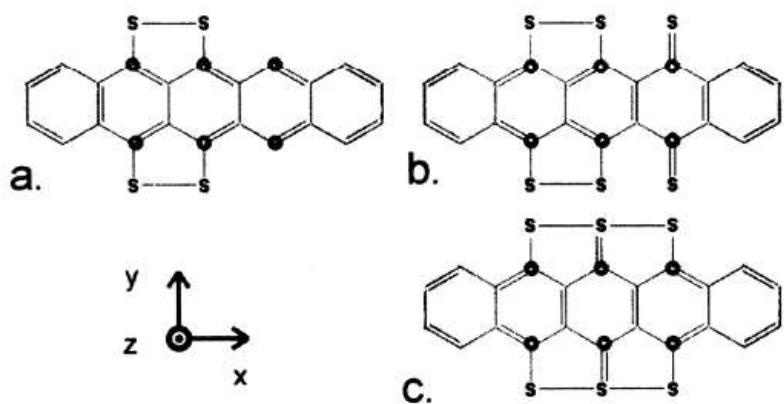

Fig. 1. Structural formulae of TTPn (a) and HTPn (b and c) molecules

After having added four S atoms, the TTPn molecule still holds two free chemically active C atoms. If S atoms form double bonds with them (thione groups >C=S), there appears the first modification of the HTPn molecule (Fig. 1,b) which can exist in two isomeric forms. In Fig. 1,c, another possible modification of the HTPn molecule, which has only one isomeric form, is displayed. By its configuration, it is close to the coronene molecule and is considered more probable. The characteristic feature of this modification is the availability of two central quadrivalent S atoms. This is possible in the case where one of the coupled $3p_x$ - electrons moves to $3d_{x^2}$-orbital.

In addition to TTPn and HTPn, other Pn thioderivatives can be formed, e.g., dithiopentacene (DTPn) with either a dithiol group or two thione groups, as in the case of tetracene thioderivatives [4].

The total number of $3p_z$-electrons in S atoms, which can participate in the conjugation with the π-system of the Pn skeleton (the meso-size effect), amounts to 10 or 8 for all HTPn or TTPn modifications, respectively. Of those electrons, the S atoms of each dithiol group supply two and the S atom of each thione group one electron. In DTPn molecules with a dithiol or two thione groups, the meso-size effect is created by four or two electrons, respectively, of the S atoms.

The absorption spectra of the Pn, PTPn, and HTPn solutions in the range 1.6–2.8 eV are depicted in Fig. 2. For a Pn solution in trichlorobenzene (curve 1), a clearly defined vibronic progression $(2.16 + 0.17n)$ eV, where $n = 0, 1, 2$, and $3$, is observed. This progression is caused by the $\pi_1 \pi_1^*$-excitation of the $2p_z$-electron which is delocalized along the perimeter of the Pn skeleton. The absorption spectrum of an HTPn solution in trichlorobenzene (curve 2) is shifted bathochromically with respect to that of Pn by 0.46 eV. The progression is less structured at that (bands at 1.70 and 1.87 eV). This progression is caused by the $l_1 \pi_1^*$-excitation of the $3p_z$ – electrons of the S atom, which interact with the π-system of the Pn skeleton [12].

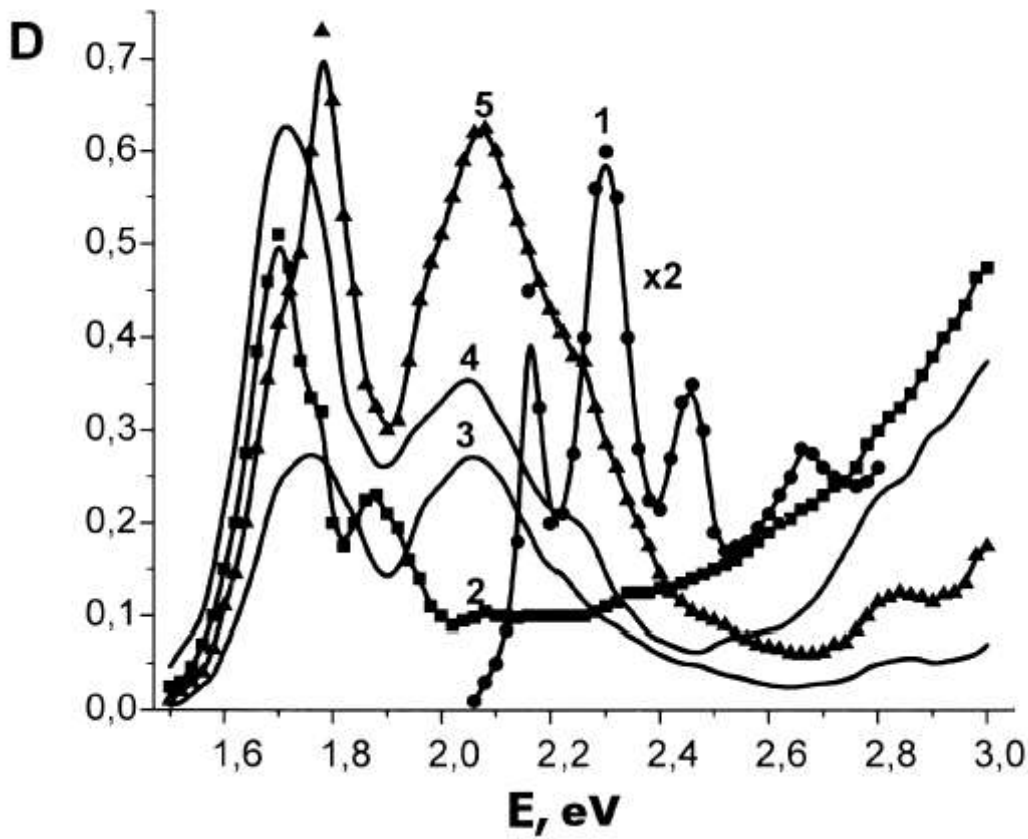

Fig. 2. Absorption spectra of Pn (curve 1, C = 4.0 × 10$^{-6}$ mol/l) and HTPn (curve 2, C = 1.0 × 10$^{-5}$ mol/l) solutions in trichlorobenzene; and of PTPn in benzene (curve 3, C = 5.4 ×10$^{-6}$ mol/l), trichlorobenzene (curve 4, C = 1.2 × 10$^{-5}$ mol/l), and DMFA (curve 5, C = 1.4 × 10$^{-5}$ mol/l)

A shoulder of 1.75-1.80 eV on the short-wave slope of the most intensive HTPn band evidences for the existence of another band. The analysis carried out in the Gaussian approximation showed that the maximum of the latter lies at 1.77 eV, as in the case of the absorption spectrum of a PTPn solution in dimethylformamide (DMFA) (curve 5 ). By position, it coincides with the $l_1 \pi_1$*-excitation bands of TTT and TSeT molecules [16].

The analysis of the long-wave absorption of PTPn solutions in benzene (curve 3 ) and trichlorobenzene (curve 4) carried out in the Gaussian approximation showed that the corresponding spectra consist of two bands each with maxima at 1.70 and 1.77 eV. Since these bands are observable for all solutions listed above, they may be considered characteristic of HTPn and TTPn, respectively. For PTPn, the relation between the peak intensities of these bands is not identical for various

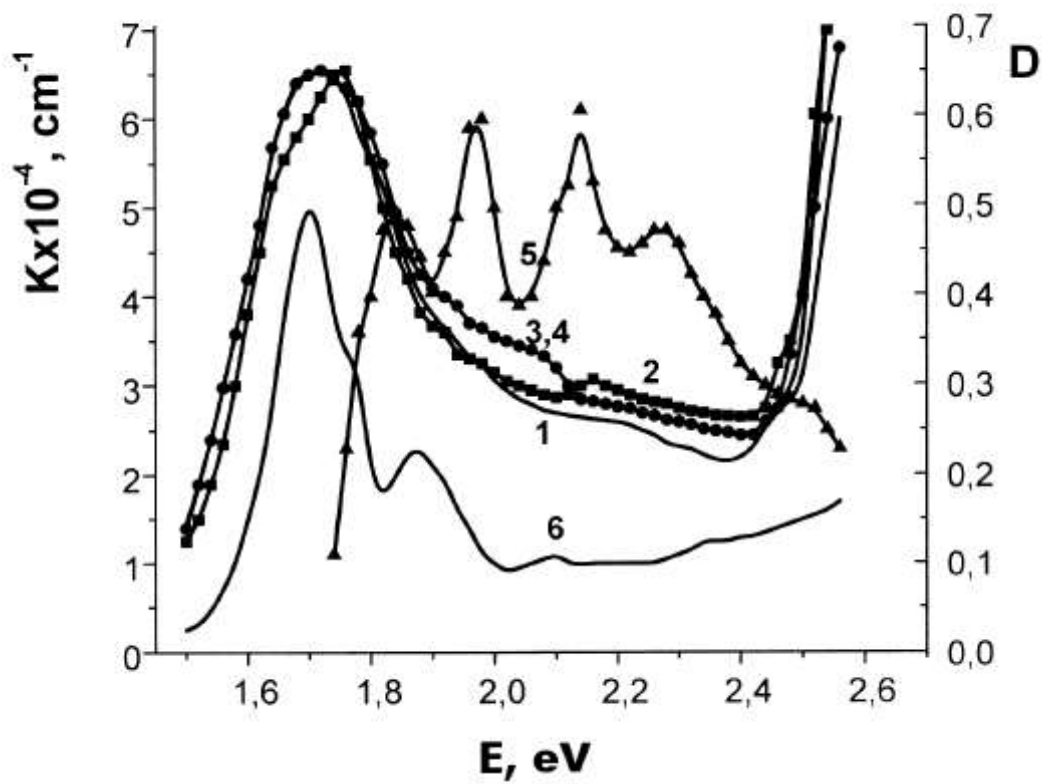

Fig. 3. Absorption spectra of HTPn (PTPn1) 100-nm films on quartz substrates (curves 1 to 4), a Pn 400-nm film (curve 5 ) [17], and a HTPn solution in trichlorobenzene (curve 6 )

solvents, which evidences for the inhomogeneity of this substance or the different solubilities of HTPn and TTPn. Moreover, a wide band at 2.07 eV is observed in PTPn spectra, being of low intensity in HTPn ones. Pentacene cannot be responsible for it, because the vibrating structure of this band is practically not pronounced, i.e. this band is caused by the excitation of the external electrons of S atoms. Most probably, this is a characteristic absorption band of DTPn with two thione groups. In this case, the bathochromic shifts of the characteristic absorption bands of HTPn, TTPn, and DTPn molecules, being accounted for one $3p_z$-electron of S atoms, are identical and equal to $(0.045 \pm 0.001)$ eV. This means that the bathochromic shift of long-wave absorption bands is described by a linear function of the number of external electrons of S atoms which take part in the conjugation. On this basis, we may assume that the 1.94-eV band is caused by the absorption of either DTPn with one dithiol or TTPn with four thione groups; it may also be a vibration duplicate of the 1.77- eV band of TTPn with two dithiol groups (coined as TTPn below).

Thus, the analysis of the absorption spectra of PTPn and HTPn solutions showed that, despite the synthesis scenario, the substances synthesized are the mixtures of Pn thioderivatives, whose main components are TTPn and HTPn. Is it possible to isolate these components? Some information can be obtained by monitoring the process of deposition of the corresponding films onto quartz substrates.

Figure 3 exposes the absorption spectra of four HTPn films, 100 nm in thickness each, in the range 1.50 – 2.50 eV. The films were deposited consistently after one another onto different substrates in the continuous process of sputtering using a mask switch. In every position of the switch, only one substrate was open for sputtering, while the other three were closed. After the first substrate having been sputtered, the second substrate was subjected to sputtering, then the third, and at last the fourth one. The analysis of the spectrum of the first deposited film (curve 1 ) showed that the contents of HTPn and TTPn in it, which were estimated by the peak intensities of the 1.66- and 1.77- eV bands, respectively, were approximately identical. In the next film (curve 2 ), TTPn prevailed. The ratios between the contents of these components in two last films (curves 3 and 4 ) were the same as in the initial one.

Therefore, the ratio between HTPn and TTPn in the films changes in the course of sputtering. Although, before sputtering, TTPn was only an impurity in the ultimately synthesized substance identified as HTPn (curve 6 and the right ordinate axis), some portion of HTPn molecules of modification b (Fig. 1) turns into TTPn ones in the course of sublimation, because they lose their thione groups. So that, in the mixture condensed on quartz substrates, the ratio between those components is close to 1:1 or TTPn prevails. We designate this mixture as PTPn1 .

It was found that the absorption spectra of PTPn and PTPn1 films are not identical in the range 1.80 – 2.00 eV. The PTPn spectral curve is located above, which evidences for a higher impurity concentration. After sublimation, the intensity of the PTPn band at 2.07 eV becomes considerably lower. This means that the third component, DTPn, decays at heating.

The bathochromic shift of the long-wave absorption bands of PTPn and PTPn1 films with respect to the spectra of the corresponding solutions amounts to 0.04 eV. It characterizes a change of the term of dispersive interaction between the molecules of these substances, when they become excited. In comparison with the Pn case, this change is five times smaller. The Davydov splitting of the bands is not observed in PTPn and PTPn1 films. For Pn crystals [17] and polycrystalline Pn films (Fig. 3, curve 5 ), the amplitude of this splitting is 0.13 eV.

There are no X-ray and electron-microscopy data for PTPn and PTPn1 films. The structure of these films can be polycrystalline, as in TTT and TSeT ones [6, 8]. In the latter substances, provided the room temperature of a substrate, round (0.1 – 0.15 μm) and needle-shaped (0.5 – 0.7 μm) crystallites are formed. The peculiarity of this structure is the availability of densely packed piles of molecules, where 3p-orbitals of the S atoms of neighbor molecules are overlapped. The calculation showed that the Davydov splitting is equal to 0.006 eV [9]. It is hard to observe this phenomenon experimentally. The contour shape of the long-wave absorption bands of TTT and TSeT films is basically determined by the deformation of dithiol groups.

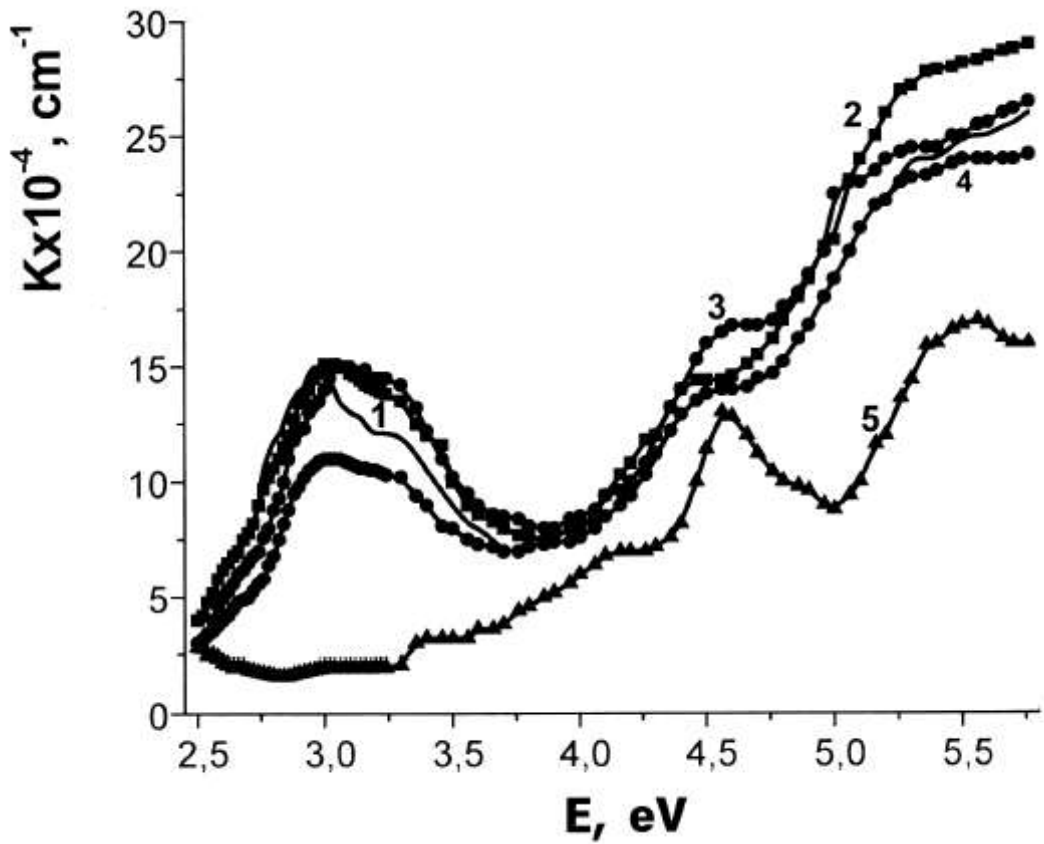

Fig. 4. Absorption spectra of HTPn (PTPn1) and Pn films. The notations are the same as in Fig. 3

In Fig. 4, the absorption spectra of Pn and PTPn1 films in the range 2.50–5.70 eV are exhibited. According to work [18], the energies of the second and third singlet transitions in Pn are 3.05 and 3.32 eV, respectively. The analysis of the absorption spectrum of the Pn film (curve 5) shows that these transitions are forbidden. One may assume that they are allowed in the spectra of PTPn1 and PTPn films. The band at 2.62 eV is observed in the absorption spectra of TTT and TSeT films. It is caused by the $l_1 \pi_1*$-excitation of $3p_z$-electrons of S atoms [16] and is the same in the spectra of PTPn and PTPn1 films. In the range 3.90 – 5.70 eV, the absorption bands of PTPn, Pn, and PTPn1 are determined by the electron transitions from lower-located occupied levels. In the PTPn and PTPn1 cases, those levels may include both $l\pi*$- and $\pi\pi*$-states.

The dark resistance of PTPn1 was measured in the cells of the surface and sandwich types. The surface cells were fabricated by the consecutive sputtering, in vacuum, of raster silver electrodes and 1-μm PTPn1 films onto quartz substrates. The distance between the electrodes was 0.2 mm. The specific resistance obtained was $10^9$ Ω × m. In sandwich cells, SnO2 and silver layers served as the lower and upper electrodes, respectively.

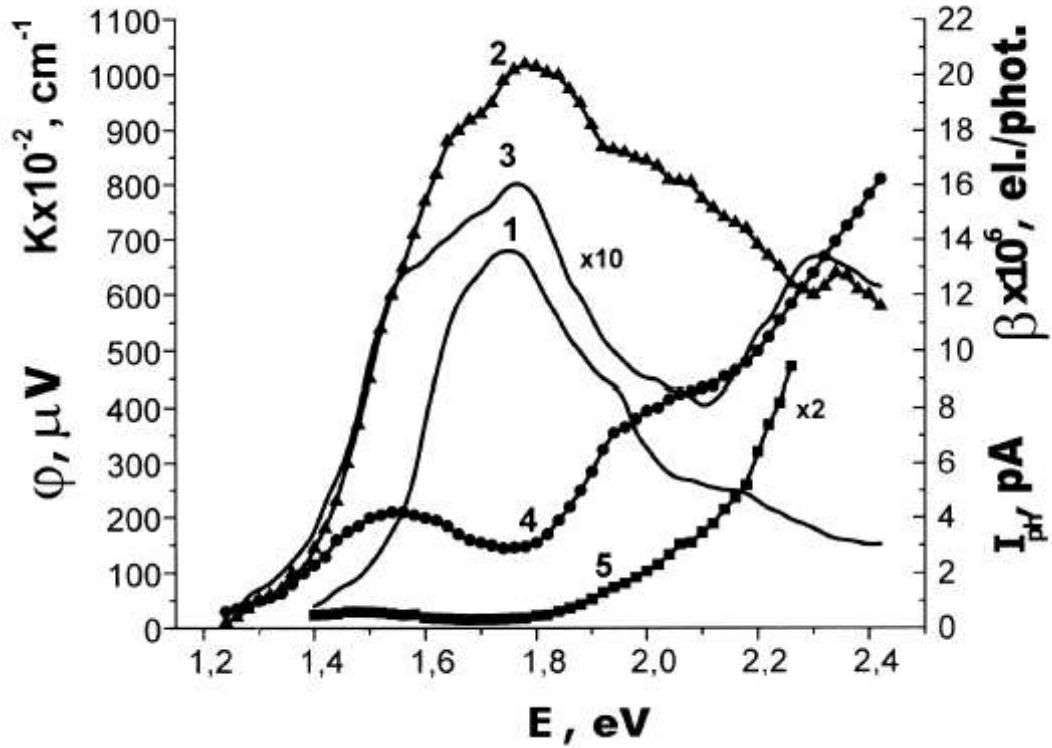

Fig. 5. Various spectra of PTPn films, 560 nm in thickness each: absorption (1); CPV ($\varphi$) under frontal (2) and back (3) modulated illumination; photocurrent $I_{ph}$ (4) and quantum efficiency $\beta$ (5) under fixed illumination

The design of the cell allowed the resistance of the silver layer to be monitored. The PTPn1 layer was found to give the main contribution to the effective resistance of the cell. The resistance of the PTPn1 layer was determined on the linear section of the current-voltage characteristics. It comprised $10^{10}$ $\Omega \times$ m, which coincides with the HTPn resistance [4]. The resistance of PTPn films was found to be of the same order as that of PTPn1 films.

The spectra of absorption, CPV, photocurrent $I_{ph}$, and quantum efficiency $\beta$ at a fixed current for PTPn films 560 nm in thickness are presented in Fig. 5. It should be emphasized that the listed spectra were measured for both PTPn and PTPn1. In this work, we presented the data only for PTPn. The results of their analysis are also valid for PTPn1.

The CPV (curve 2) and absorption (curve 1) spectra correlate in the range 1.30 – 2.55 eV, when illuminating the free surface of PTPn through the frontal SnO2 electrode and the Teflon film. In so doing, the free surface becomes charged negatively, while the electrode, owing to electrostatic induction, positively. The illumination through the back electrode does not change the CPV sign, but makes its peak intensities an order of magnitude lower in comparison with those for the frontal CPV (curve 3). The analysis showed that the long-wave structural portion of the back CPV spectrum consists of three bands with maxima at 1.54, 1.68, and 1.77 eV. The 1.54-eV band, by its position, coincides with that in the photocurrent spectrum (curve 4, the right ordinate axis); the latter anticorrelates with the absorption spectrum of PTPn.

The magnitude of the quantum efficiency β(E) (curve 5, the right ordinate axis) was estimated by the formula

$$\beta(E) = I_{ph} E / eP(1 - \exp(-kd)), \quad (1)$$

where $I_{ph}$ is the photocurrent, E the energy of photons, $e = 1.6 \times 10^{-19}$ C, P is the light flux power, k the absorption coefficient, and d the thickness of PTPn films. Formula (1) does not take the reflection coefficient of PTPn films into account, so that the obtained values of β(E) are the minimal possible ones.

In the range of the first singlet transition in PTPn, β(E) amounts to $10^{-7}$ electron/photon. With increase in the photon energy, β(E) grows monotonously and becomes equal to $5 \times 10^{-6}$ electron/photon at E = 2.26 eV, i.e. increases by more than an order of magnitude. In this case, the ratio of the photo- and dark currents is equal to 1.6 in an electric field of 250 kV/m.

The dependence of the CPV amplitude on the wavelength of exciting light λ can be expressed by the equation

$$\varphi = Ak / (1 + kL), \quad (2)$$

where A is a constant, and the absorption coefficient k is unequivocally connected with λ [13].

The extrapolation of the dependence $\varphi^{-1}(k^{-1})$, which had been measured in the range of the first electron transition, up to its intersection with the abscissa axis allowed us to estimate the diffusion length of singlet excitons L = (210 ± 50) nm for a PTPn layer 560 nm in thickness, which agrees well with the data of work [13].

After air had been pumped out of the cell, the frontal CPV increased within the whole spectral range under study, not violating the correlation with the absorption spectrum. In vacuum of 13.5 mPa, the increase of PTPn absorption in its maximum (at E = 1.77 eV) amounted to 9%. At letting-to-air, the CPV diminished to the initial value level.

The CPV is considered as the algebraic sum of the bulk (Dember) photovoltage (PV) $\varphi_D$, the surface-barrier PV $\varphi_b$, and the surface PV $\varphi_c$ emerging owing to the capture of nonequilibrium charge carriers by surface states of the donor or acceptor type [19]. The bulk $\varphi_D$ arises under the condition of bipolar diffusion, due to the difference between the mobilities of charge carriers with different signs. In the case of unipolar photoconductivity, it is governed by the gradient of the concentration of nonequilibrium charge carriers. The surface-barrier PV $\varphi_b$ appears owing to the separation of free charge carriers by the electric field of the space-charge region (SCR). Its sign is defined by the direction of the band bending under equilibrium conditions.

Therefore, according to the experimental data presented above, the illuminated free surface of PTPn becomes charged negatively. This evidences for a higher mobility of holes in the case of bipolar photoconductivity or for their diffusion into the bulk in the case of unipolar photoconductivity. In the range of the first singlet transition in PTPn, unipolar photoconductivity is more probable, which is confirmed by the spectral dependence β(E). The sign of the frontal CPV testifies to the hole character of this photoconductivity which is caused by the ionization of excitons at electron capture centers.

The overlapping of 3p-orbitals of the S atoms of neighboring molecules stimulates the appearance of linear chains in PTPn and PTPn1 crystallites; the chains being arranged in parallel to the axes of regular molecular piles, which are composed of translationally equivalent molecules. Probably, excitons may diffuse along these one-dimensional chains.

The sign and the amplitude of $\varphi_D$ depend on the concentration gradient of nonequilibrium charge carriers, while the sign and the amplitude of $\varphi_b$ on their concentration and the drift direction in the electric field of the SCR near the PTPn free surface. Accordingly, the spectra of $\varphi_D$ and $\varphi_b$ have to be close to the absorption or the photocurrent one, respectively [20]. The frontal CPV correlates with the absorption spectrum, i.e. the main contribution is made here by $\varphi_D$ ($\varphi_D \gg \varphi_b$). The 1.54-eV band in the back CPV spectrum evidences for its drift character, and the CPV sign for the antilocking band bending for holes near the back electrode. Since the back CPV is much lower by amplitude than the frontal one and does not change its sign in the range of strong PTPn absorption, it testifies to that the bands at 1.68 and 1.77 eV are caused by $\varphi_b$ ($\varphi_b > \varphi_D$). Moreover, the photogeneration of holes near the back electrode is caused by the dissociation of excitons owing to the capture of electrons onto acceptor levels with various energies.

The increase of the frontal CPV in vacuum can be explained by a reduction of the concentration of oxygen molecules, which are electron acceptors and are adsorbed on the free PTPn surface. In this case, the antilocking band bending decreases, and the CPV amplitude ($|\varphi_D - \varphi_b|$) increases.

4. Conclusions

Mixtures, which are called polythiopentacenes, are formed in all scenarios of the synthesis of Pn thioderivatives. After their condensation onto quartz substrates, their main components are TTPn and HTPn, which we did not manage to isolate in the course of sputtering.

In the spectral range 1.30 – 2.40 eV, the photoconductivity of PTPn films is of the hole type and is caused by the dissociation of excitons owing to their collisions with electron capture centers. The frontal CPV in this range is caused by the value of $\varphi_D$, while the back one by $\varphi_b$.

1. Marshalk C. // Bull. Soc. Chim. France. 1939. 6. P. 1122.
2. Marshalk C., Stumm C. // Ibid. 1948. 15. P. 418 - 429.
3. Matsunaga Y. // J. Chem. Phys. 1965. 42, N 6. P. 2248 - 2249.
4. Goodings E.P., Mitchard D.A., Owen G. // J. Chem. Soc. Perkin Trans. I. 1972. N 11. P. 1310 - 1314.
5. Balode D.R., Silinsh E.A., Taure L.F. // Izv. AN Latv. SSR. Ser. Fiz. Tekhn. Nauk. 1978. N 1. P. 35 - 45.
6. Vertsimakha Ya.I., Gorishnyi M.P., Kurik M.V., Libera L. // Ukr. Fiz. Zh. 1981. 26, N 10. P. 1704 - 1709.
7. Perez-Albuerne E.A., Johnson H., Trevoy D.J. // J. Chem. Phys. 1971. 55, N 4. P. 1547 - 1554.
8. Shibaeva R.P. // Totals of Science and Technique. Ser. Crystallochem. / Ed. by. E.A. Gilinskaya. Moscow: VINITI, 1981 (in Russian). P. 189 - 264.
9. Gorishnyi M.P. // Abstract of Ph.D. thesis "Electronic Structure of Tetrathiotetracene and Photoelectric Properties of Heterostructures on Its Basis". Lviv, 1990 (in Russian).
10. Vertsimakha Ya.I. // Abstract of Ph.D. thesis "Mechanisms of Charge Carrier Photogeneration and Energy Structure of Pentacene". Kyiv, 1978 (in Russian).
11. Verzimacha J.I., Gorišnyj M.P., Kurik M.V., Libera L. // 19. Tagung "Organische Festk¨orper". Erfurt, 1982. S. 142 - 144.
12. Gorishnyi M.P., Kurik M.V., Libera L. // Ukr. Fiz. Zh. 1987. 32, N 7. P. 1013 - 1016.
13. Lutsyk P., Vertsimakha Ya. Organic p-n type heterostructures based on the hexatiopentacene // Mol. Cryst. and Liquid Cryst. 2005. 426. P. 265 - 276.
14. Ugawa A., Kunikiyo T., Ohta Y. et al. // MAR05 Meeting of the American Phys. Soc., 2004.
15. Meshkov A.M., Akimov I.A. // Prib. Tekhn. Eksp. 1964. N 3. P. 181 - 185.
16. Gorishnyi M.P. // Ukr. Fiz. Zh. 2002. 47, N 8. P. 711 - 714.
17. Silinsh E.A., Kurik M.V., Chapek V. Electronic Processes in Organic Molecular Crystals. Riga: Zinatne, 1988 (in Russian).
18. Silinsh E.A., Belkind A.I., Balode D.R. et al. // Phys. status solidi (a). 1974. 25. P. 339 - 347.
19. Peka G.P., Strikha V.I. Surface and Contact Phenomena in Semiconductors. Kyiv: Lybid, 1992 (in Ukrainian).
20. Akimov I.A. // Opt. Mekh. Prom. 1966. N 5. P. 4 - 13.

МОЛЕКУЛЯРНА СТРУКТУРА ТА ЕЛЕКТРОФІЗИЧНІ ВЛАСТИВОСТІ ТІОПОХІДНИХ ПЕНТАЦЕНУ

М.П. Горішний

Р е з ю м е

Розглянуто можливі конфігурації молекул тіопохідних пентацену. Досліджено спектри поглинання розчинів і плівок політіопентацену (ПТП). Встановлено, що ПТП це суміш тіопохідних пентацену з різним числом атомів S. Після конденсації цієї суміші у вакуумі на кварцові підкладки головними її компонентами є тетратіопентацен (ТТП) і гексатіопентацен (ГТП). Положення максимуму довгохвильових смуг поглинання тіопохідних пентацену описується лінійною функ- цією числа валентних електронів атомів S, які беруть участь у спряженні із π-системою пентаценового каркаса молекул ПТП. Аналіз спектрів фотоструму і конденсаторної фото-ерс (КФЕ) в області перших електронних переходів ПТП показав, що фо- топровідність є дірковою і зумовлена дисоціацією екситонів на центрах захоплення електронів. Фронтальна КФЕ зумовле- на дифузійною фото-ерс. Дембера φД , тильна поверхнево- бар'єрною фото-ерс φб .